\documentstyle [12pt] {article}

\def\L{{\cal L}}
\def\R{{\rm I \!\!\, R}}
\def\be{\begin{equation}}
\def\ee{\end{equation}}
\def\bea{\begin{eqnarray}}
\def\eea{\end{eqnarray}}

\begin{document}
\baselineskip=24pt

\begin{titlepage}
\title{ {\bf SPACE-TIMES ADMITTING A THREE-DIMENSIONAL CONFORMAL GROUP}}
\author{{\normalsize J. Carot}\thanks{Departament de F\'{\i}sica,
Universitat de les Illes Balears,  E-07071 Palma de Mallorca. SPAIN.}\and
{\normalsize A.A. Coley}\thanks{ Department of Mathematics, Statistics and
Computing Science,  Dalhousie University, Halifax, Nova Scotia. CANADA B3H
3J5.} \and {\normalsize  A.M. Sintes*}}
 \date{}
 \maketitle
\end{titlepage}


\centerline{ \Large{ \bf Abstract}}
\vskip1cm
Perfect fluid space-times admitting a three-dimensional Lie group of
conformal motions containing a two-dimensional Abelian Lie subgroup of
isometries are studied. Demanding that the conformal Killing vector be
proper (i.e., not
homothetic nor Killing), all such space-times are classified according to
the structure of their corresponding three-dimensional conformal Lie group
and the
nature of their corresponding orbits (that are assumed to be non-null).
Each metric is then explicitly displayed in coordinates adapted to the
symmetry vectors. Attention is then restricted to
the diagonal case, and exact perfect fluid solutions are obtained in both
the cases in which the fluid four-velocity is tangential or orthogonal to the
conformal orbits, as well as in the more general \lq\lq tilting" case.

\vskip2cm


 \newpage

\section{Introduction}
\indent We shall study perfect fluid space-times admitting a
three-dimensional
Lie group of conformal motions $C_3$, which contains a two-dimensional
Abelian Lie subgroup of isometries $G_2$.
The precise assumptions are:\begin{enumerate}
\item The $G_2$ acts orthogonally transitively on two-dimensional spacelike
orbits diffeomorphic to $\R^2$.
\item One of the Killing vectors (KV) in $G_2$ is hypersurface orthogonal, hence
the metric can be written in diagonal form \cite{Hewitt88}.
\item The conformal Killing vector (CKV) is proper (non-homothetic).
\item The orbits of the $C_3$ are non-null. \end{enumerate}
Space-times of this type have been studied recently by Kramer and Carot
\cite{KrCar} and Castej\'{o}n-Amenedo and Coley \cite{CACol}. Here we shall
systematically classify geometrically all possible space-times according to
their (inequivalent) group structures and the nature of their corresponding
orbits, and thereafter attempt to obtain the perfect fluid models in each
class. In this classification scheme the (stationary and axisymmetric)
models
studied by  Kramer \cite{Kr90} and Kramer and Carot \cite{KrCar} admit a
$C_3$
acting on three-dimensional timelike hypersurfaces. The models studied by
Castej\'{o}n-Amenedo and Coley  \cite{CACol} admit a $C_3$ acting on
two-dimensional timelike hypersurfaces containing an Abelian $C_2$. Mars and
Senovilla are looking at models in which $C_3$ is Abelian and
trying to extend their study to other non-diagonal cases \cite{MarSen93} and
Coley and Czapor \cite{ColCza93} are studying  space-times in which the
CKV is inheriting \cite{ColTup90a}.

This work generalizes previous work in several ways. First, solutions
admitting  a CKV in space-times with additional symmetry have been studied;
for example, spatially homogeneous \cite{ColTup90a}, spherically symmetric
(see, for instance \cite{ColTup90b} and references cited therein) and plane
symmetric \cite{ColCza92} models have been investigated. In a sense the
present work  is the natural generalization of this research in that  CKV are
studied in space-times with the next highest degree of symmetry. Indeed, it is
for this reason that $G_2$ space-times have begun to attract much attention.
Cosmological models with an Abelian $G_2$ acting on  spacelike hypersurfaces
have been studied by Hewitt et al.\cite{Hewitt88}, Hewitt and Wainwright
\cite{Hewitt90}, Hewitt et al.\cite{Hewitt91}, Ruiz and Senovilla \cite{Ruiz}
and Van den Berg and Skea \cite{Bergh92}. Models with an Abelian $G_2$ acting
on timelike hypersurfaces, including the astrophysically relevant stationary
and axisymmetric models, which have been studied for many years \cite{Kramer},
have also attracted renewed attention
\cite{Kramer84,Kramer90,Senovilla92,Mars}.
Second, the cases in which the CKV degenerates to either a KV or a homothetic
vector field (HVF) have been studied previously. Space-times admitting a $G_3$
have received much attention \cite{Kramer}. Space-times admitting a HVF in
addition to two KV have been studied by Hewitt and Wainwright \cite{Hewitt90}
and Carot et al. \cite{Ali1}.
Third, this work generalizes that of Coley and Czapor \cite{ColCza93} in
which the CKV is inheriting, and complements that of Senovilla
\cite{Senovilla92} in which the Abelian $G_2$ acts on two-dimensional timelike
hypersurfaces (these models are, in fact, stationary and axisymmetric perfect
fluid solutions admitting a proper CKV).
Finally, we shall assume that the perfect fluid matter  satisfies $\mu>0$ (all
vacuum space-times admitting a proper CKV are known,
\cite{ColTup90a,Hall90c}).\hfill\break

For a very precise and accurate study of the general properties of CKVs,
their Lie algebra and fixed point structure, we refer the reader to
\cite{Hall90b}.\hfill\break

The outline of this paper is as follows. In section 2 we shall describe the
models under consideration in detail, defining all relevant terms. We shall
then classify the space-times into  different inequivalent classes, in each
of the two cases in which the CKV is  everywhere spacelike or not (i.e., the
CKV can be timelike, see Eqns. (\ref{c9}) and (\ref{c10})),
according to the structure of their corresponding three-dimensional conformal
Lie group $C_3$ (each containing an Abelian $G_2$). We shall display the
corresponding metrics (and symmetry vectors) in coordinates adapted to the
CKV. These results are completely independent of the Einstein field equations
and the assumed energy-momentum tensor. We note parenthetically that not all
the classes may be possible (again, irrespective of the assumed matter
content) in some situations of physical interest; for example, Mars and
Senovilla \cite{Mars} have shown that in axially symmetric space-times
admitting a maximal three-dimensional (or two-dimensional) conformal Lie
algebra (such as in a cylindrically symmetric space-time with one CKV), the
axial KV must commute with the other two (in fact, they show that this is true
regardless of whether the additional KV is spacelike, such as, for example,
in the case of a
stationary and axially symmetric space-time with one CKV).

In section 3 we study those cases where one KV is hypersurface orthogonal,
hence the metric can be written in diagonal  form.
In sections 4 and 5 we then utilize Einstein's field equations in an attempt
to find perfect fluid solutions representative of the classes in each of
the two
cases corresponding to the form of the CKV. We shall assume that the matter
satisfies  the weak and dominant energy
conditions ($\mu>0$, $-\mu\leq p\leq\mu$). In the final section all of the
perfect fluid solutions obtained are summarized and briefly discussed.

\section{Space-times admitting conformal Killing vectors}

As we have already pointed out, we shall concern ourselves with
space-times $(M,g)$
that admit a three-parameter conformal group $C_3$ containing an Abelian
two-parameter subgroup of isometries $G_2$, whose orbits $S_2$ are spacelike,
diffeomorphic to $\R^2$ and admit orthogonal two-surfaces;
furthermore, we shall assume that the $C_3$  acts transitively on non-null
orbits $V_3$. As is customary, we shall denote the KV spanning ${\cal G}_2$
(the Lie algebra associated with $G_2$) by $\xi$ and $\eta$, and the proper
CKV in
${\cal C}_3$ (the Lie algebra of the conformal group $C_3$) by $X$. Since, by
hypothesis, $\xi$ and $\eta$ commute, we can (locally) adapt two coordinates,
say $y$ and $z$, so that
\begin{equation}
\xi={\partial\over\partial y}\ ,\qquad \eta={\partial\over\partial z}\ .
\end{equation}
Taking now two more coordinates $t$ and $r$, it follows from the above
assumptions that the line element associated with the metric $g$ can be written
as \cite{Wain}
\begin{equation}
ds^2=e^{2F}\{-dt^2+dr^2+Q[h^{-2}(dy+wdz)^2+h^2dz^2]\}\ ,
\label{c2}
\end{equation}
where $F$, $Q$, $h$ and $w$ are all functions of $t$ and $r$ alone.

The (proper) CKV $X$ will satisfy
\begin{equation}
(\L_Xg)_{ab}=2\Psi g_{ab}\ ,
\label{c3}
\end{equation}
where $\L$ stands for the Lie derivative operator and $\Psi=\Psi(x^c)$ is the
so-called conformal factor (the particular cases $\Psi=0$ and
$\Psi=const\not=0$ correspond, respectively, to $X$ being a KV and a
proper HVF). Assuming that no further CKV exist on $M$ (i.e., the $C_3$ is
maximal) one has the following two families of Lie algebra structures for
${\cal C}_3$:
\bea
&({\rm A})&\quad[\xi,\eta]=0\ ,\quad [\xi,X]=\alpha_1 \xi+\alpha_2\eta \ ,\quad
[\eta,X]=\beta_1\xi+\beta_2\eta\ ,\\
&({\rm B})&\quad[\xi,\eta]=0\ , \quad [\xi,X]=a_1 \xi+a_2\eta +a_3X\ , \quad
[\eta,X]=b_1\xi+b_2\eta+b_3X\ ,
\eea
where the $\alpha_i$, $\beta_i$, $a_i$ and $b_i$ are constants.

The family ${\cal C}_3({\rm A})$ can, in turn, be classified into the following
seven Bianchi types (see for instance \cite{Petrov}):
\begin{eqnarray}
(I)&\quad &[\xi,\eta]=[\xi,X]=[\eta,X]=0\ , \nonumber \\
(II)&\quad &[\xi,\eta]=[\xi,X]=0 \quad [\eta,X]=\xi\ ,  \nonumber \\
(III)&\quad &[\xi,\eta]=0\quad [\xi,X]=\xi \quad [\eta,X]=0\ , \nonumber \\
(IV)&\quad &[\xi,\eta]=0\quad [\xi,X]=\xi \quad [\eta,X]=\xi+\eta ,\label{c6}
\\ (V)&\quad &[\xi,\eta]=0\quad [\xi,X]=\xi \quad [\eta,X]=\eta\ ,  \nonumber \\
(VI)&\quad &[\xi,\eta]=0\quad [\xi,X]=\xi \quad [\eta,X]=q\eta \quad (q\not=
0,1)\ ,  \nonumber \\
(VII)&\quad &[\xi,\eta]=0\quad [\xi,X]=\eta \quad
[\eta,X]=-\xi+q\eta \quad  (q^2<4)\ , \nonumber
\end{eqnarray}
whereas family ${\cal C}_3$(B) can be brought to the following form by
means of appropriate
re-definitions of $X$ and the KV's $\xi$ and $\eta$:
\be
[\xi,\eta]=0\quad [\xi,X]=X\quad [\eta,X]=0\ .
\label{c7}
\ee
Notice that in the case of $X$ being a HVF the above algebraic structure
(\ref{c7}) is forbidden, since the Lie bracket of a HVF and a KV must
necessarily be a KV (for further information on this case, see \cite{Ali1}).

Assuming now for the CKV $X$ an expression of the form
\be
X=X^a(x^c)\partial_a\ ,
\label{c8}
\ee
in the coordinate chart $\{t,r,y,z\}$, and specializing the equation (\ref{c3})
to the metric given in (\ref{c2}) and the CKV given in (\ref{c8}), it is easy
to see that one can always, by means of a coordinate transformation in the
$t,r$ plane, bring $X$ to one of the following forms:
\bea
(a)&\,& X=\partial_t+X^y(y,z)\partial_y+X^z(y,z)\partial_z\ , \nonumber\\
(b)&\,& X=\partial_r+X^y(y,z)\partial_y+X^z(y,z)\partial_z\ , \label{c9} \\
(c)&\,& X=\partial_t+\partial_r+X^y(y,z)\partial_y+X^z(y,z)\partial_z\ ,
\nonumber \eea
if the conformal algebra ${\cal C}_3$ belongs to the family (A), whence
$X^y(y,z)$ and $X^z(y,z)$ are linear functions of their arguments to be
determined from the commutation relations of $X$ with $\xi$  and $\eta$ (see
(\ref{c6}-$(I)$) to (\ref{c6}-$(VII)$); and to the forms:
\bea
(a)&\,& X=e^y(\partial_t+X^y(t)\partial_y+X^z(t)\partial_z)\ , \nonumber\\
(b)&\,& X=e^y(\partial_r+X^y(r)\partial_y+X^z(r)\partial_z)\ , \label{c10}\\
(c)&\,& X=e^y(\partial_t+\partial_r+X^y(t,r)\partial_y+X^z(t,r)\partial_z)\ ,
\nonumber \eea
if ${\cal C}_3$  is that given by (\ref{c7}) (family (B)).

The forms (\ref{c9}-$a$) and (\ref{c10}-$a$) are easily seen to correspond to
the case of three-dimensional timelike conformal orbits $T_3$, whereas
(\ref{c9}-$b$) and (\ref{c10}-$b$) correspond to  three-dimensional spacelike
conformal orbits $S_3$. The remaining possibilities, (\ref{c9}-$c$) and
(\ref{c10}-$c$), imply null conformal orbits  $N_3$ and will not be
considered in the present paper.

Assuming now the form (\ref{c9}-$a$) for the CKV $X$ (i.e., family (A),
timelike conformal orbits), one then has for each possible case ($I$) to
($VII$) the following forms for $X$ and the metric functions:
\bea
(I)&\,& Q=\hat Q(r)\ , \quad h^2=\hat h^2(r)\ , \quad w=\hat w(r)\ , \nonumber\\
&\,& X=\partial_t\ . \label{c11}\\
(II)&\,& Q=\hat Q(r)\ , \quad h^2=\hat h^2(r)\ , \quad w=\hat w(r)-t\ ,
\nonumber\\
&\,& X=\partial_t+z\partial_y\ . \label{c12}\\
(III)&\,& Q=e^{-t}\hat Q(r)\ , \quad h^2=e^t\hat h^2(r)\ , \quad w=e^t\hat
w(r)\ , \nonumber\\
&\,& X=\partial_t+y\partial_y\ . \label{c13}\\
(IV)&\,& Q=e^{-2t}\hat Q(r)\ , \quad h^2=\hat h^2(r)\ , \quad w=\hat w(r)-t\ ,
\nonumber\\
&\,& X=\partial_t+(y+z)\partial_y+z\partial_z\ .\label{c14} \\
(V)&\,& Q=e^{-2t}\hat Q(r)\ , \quad h^2=\hat h^2(r)\ , \quad w=\hat w(r)\ ,
\nonumber\\ &\,& X=\partial_t+y\partial_y+z\partial_z\ .\label{c15} \\
(VI)&\,& Q=e^{-(1+q)t}\hat Q(r)\ , \quad h^2=e^{(1-q)t}\hat h^2(r)\ , \quad
w=e^{(1-q)t}\hat w(r)\ , \nonumber\\ &\,&
X=\partial_t+y\partial_y+qz\partial_z\quad (q\not=0,1)\ .\label{c16} \\
(VII)&\,& Q=e^{-qt} {\sqrt{4-q^2}\over 2} a(r)\ ,  \nonumber\\
&\,& h^2={{\sqrt{4-q^2}\over 2}a(r)\over \sqrt{a(r)^2+c(r)^2+g(r)^2}
+c(r)\cos(\sqrt{4-q^2}t) +g(r)\sin(\sqrt{4-q^2}t)}\ , \nonumber\\
&\,& w={q\over 2}+{{\sqrt{4-q^2}\over 2}
[c(r)\sin(\sqrt{4-q^2}t)-g(r)\cos(\sqrt{4-q^2}t)] \over
\sqrt{a(r)^2+c(r)^2+g(r)^2} +c(r)\cos(\sqrt{4-q^2}t)
+g(r)\sin(\sqrt{4-q^2}t)}\ , \nonumber\\
&\,&
X=\partial_t-z\partial_y+(y+qz)\partial_z\quad (q^2<4)\ . \eea

In all of these cases $F=F(t,r)$ and the conformal factor, $\Psi$, is given by
\be
\Psi=F_{,t}\label{c18}
\ee

The form (\ref{c9}-$b$) for the CKV $X$ (i.e., family A, spacelike conformal
orbits) would yield similar results with the role of the coordinates $t$ and
$r$ reversed.

Similarly, if we take the form (\ref{c10}-$a$) for $X$ (Family B, timelike
conformal orbits) we get, assuming the canonical form (\ref{c7}) for ${\cal
C}_3$, the following possibilities for the metric and the CKV:
\bea
(1)&\,& ds^2=e^{2F}\{-dt^2+dr^2+[\Phi^2(r)(c\cosh t+d)^2+\sinh^2t]dy^2
 +\nonumber\\ &\,&
+2\Phi^2(r)(c\cosh t+d)dydz+\Phi^2(r)dz^2\}\ , \nonumber\\
&\,& X=e^y\left(\partial_t-\coth t\partial_y+{c+d\cosh t\over \sinh
t}\partial_z\right)\ , \label{c19}\\
(2)&\,& ds^2=e^{2F}\{-dt^2+dr^2+[\Phi^2(r)(c t^2+d)^2+t^2]dy^2 +\nonumber\\
&\,&
+2\Phi^2(r)(c t^2+d)dydz+\Phi^2(r)dz^2\}\ , \nonumber\\
&\,& X=e^y\left(\partial_t-{1\over t}\partial_y-{ct^2-d\over \vert
t\vert}\partial_z\right)\ ,\label{c20}\\
(3)&\,& ds^2=e^{2F}\{-dt^2+dr^2+[\Phi^2(r)(d-c\cos t)^2+\sin^2t]dy^2 +
\nonumber\\ &\,&
+2\Phi^2(r)(d-c\cos t)dydz+\Phi^2(r)dz^2\}\ , \nonumber\\
&\,& X=e^y\left(\partial_t-\cot t\partial_y+{d\cos t-c\over \sin t}
\partial_z\right)\ ,\label{c21} \eea
where $c$ and $d$ are constants, $F=F(t,r)$ is an arbitrary function of $t$
and $r$, and the conformal factor $\Psi$ is given by
\be
\Psi=e^yF_{,t}\ .\label{c22}
\ee

The case of spacelike conformal orbits for the family B (i.e., $X$ of the form
(\ref{c10}-$b$)) leads to just one possibility, namely
\bea
&\,& ds^2=e^{2F}\{-dt^2+dr^2+[\Phi^2(t)(c\sinh r+d)^2+\cosh^2r]dy^2+
\nonumber\\ &\,&
+2\Phi^2(t)(c\sinh r+d)dydz+\Phi^2(t)dz^2\}\ , \nonumber\\
&\,& X=e^y\left(\partial_r-\tanh r\partial_y+{c-d\sinh r\over \cosh
r}\partial_z\right)\ ,
\eea
where $c$ and $d$ are again constants, $F=F(t,r)$ is an arbitrary function of
its variables and the conformal factor is in this case
\be
\Psi=e^yF_{,r}\ .
\ee

\section{Diagonal Case.}
\label{se3}
Next we will study those cases where two hypersurface orthogonal KVs exist
in ${\cal G}_2$; this implies that the two KVs must be mutually orthogonal and
therefore, by means of a linear change of coordinates in the Killing orbits
$S_2$, one can always set $w$ in (\ref{c2}) to zero and the metric then becomes
diagonal \cite{Wain} .

It is worth noticing that this is not possible for types $II$ and $IV$ in
family A, as one can see by simply inspecting the form of $w$ in these  cases
(see (\ref{c12}) and (\ref{c14})). As for the case $VII$ (also belonging to
family A), $w=0$ implies the existence of a third KV tangent to the Killing
orbits $S_2$, in which case they are of constant curvature and the conformal
algebra becomes four dimensional; therefore, we shall not consider this case
further.
For those metrics in family B we have:
\bea
 ds^2 &= & e^{2F}\{-dt^2+dr^2+\sinh^2tdy^2  + \Phi^2(r)dz^2\}, \nonumber\\
 X &=&  e^y\left(\partial_t-\coth t\partial_y\right),\label{c25}\\
 ds^2 &=& e^{2F}\{-dt^2+dr^2+t^2dy^2 +\Phi^2(r)dz^2\}, \nonumber\\
X &= &e^y\left(\partial_t-{1\over t}\partial_y \right),\label{c26}\\
ds^2 &= &e^{2F}\{-dt^2+dr^2+\sin^2tdy^2 +\Phi^2(r)dz^2\}, \nonumber\\
X &=& e^y\left(\partial_t-\cot t\partial_y\right),\label{c27} \eea
if the conformal orbits are timelike, and, for spacelike conformal orbits:
\bea
ds^2 &= &e^{2F}\{-dt^2+dr^2+\cosh^2rdy^2 +\Phi^2(t)dz^2\}, \nonumber\\
X& =& e^y\left(\partial_r-\tanh r\partial_y\right).
\eea
There is still another possibility for metrics belonging to this family;
namely,
\bea
 ds^2 &=& e^{2F}\{-dt^2+dr^2+\cos^2tdy^2 +\sin^2t dz^2\}, \nonumber\\
X &= &e^{y+z}\left(\partial_t+\tan t\partial_y -\cot t\partial_z\right),
\eea
but   this metric admits a further proper CKV, $\partial_r$,
therefore the maximal conformal group is four-dimensional and again we shall
not consider this case any further.

\section{Perfect fluid space-times: Fluid flow tangent or orthogonal to the
conformal orbits.}
\label{se4}
In this section we will study diagonal perfect fluid solutions admitting a
maximal $C_3$ of conformal motions. The energy-momentum tensor for a perfect
fluid is given by
\be
T_{ab}= (\mu+p) u_au_b+ p g_{ab}\ ,\label{c29}
\ee
where $\mu$ and $p$ are, respectively, the energy density and the pressure
as measured by
observers comoving with the fluid, and $u^a$ ($u^au_a=-1$) is the
four-velocity of the fluid. Since the metric admits two KVs, $\xi$  and
$\eta$, it follows that the Lie derivative   of $p$, $\mu$ and
$u_a$ with respect to each must vanish identically; in our coordinate chart
this is equivalent to:
\be
\mu=\mu(t,r) \quad p=p(t,r) \quad u_a=u_a(t,r)\ .
\ee
Since the isometry orbits admit orthogonal two-surfaces one has that, in these
coordinates, both the metric and the Ricci tensor have block-diagonal forms;
thus, the Einstein  field equations specialized to a perfect fluid, as
given by (\ref{c29}), reduce simply to
\be
u_y=u_z=0\ ,
\ee
\be
{G_{yy}\over g_{yy}}-{G_{zz}\over g_{zz}}=0\ ,\label{c32}
\ee
\be
G_{tr}^2-\left(G_{tt}-{G_{yy}\over g_{yy}}g_{tt}\right)
\left(G_{rr}-{G_{yy}\over g_{yy}}g_{rr}\right)=0 \label{c33}
\ee
(where we have already taken into account that the metric is diagonal,
otherwise we would have had an extra equation, namely: $G_{yy}/
g_{yy}-G_{yz}/ g_{yz}=0$).

We shall focus on those solutions such that the perfect fluid four-velocity is
either tangent or orthogonal to the conformal orbits; the general case (or
\lq\lq tilted" case) will be the subject of the next section.

\subsection{Fluid flow tangent to the conformal orbits.}

This case can only arise when the conformal orbits are timelike, and then one
has:
\be
u_t=-e^F \quad u_r=0
\ee
(i.e., the coordinates are comoving). The metric must be one of those given by
(\ref{c11}), (\ref{c13}), (\ref{c15}) and (\ref{c16}) (with $w=0$) if belongs
to family A (Bianchi types $I$, $III$, $V$ and $VI$, respectively), or one of
those given by (\ref{c25}), (\ref{c26}) and (\ref{c27}) if it belongs to family
B.

We next summarize the results obtained under these hypotheses for metrics of
both
families A and B.\hfill\break

{\bf Family A:}\hfill\break

Type $I$: In this case, the fluid velocity is parallel to the CKV $X$. This
case has  already been studied in full generality  (see \cite{Coley91}) showing
that the only such space-times are locally FRW models (assuming $\mu+p\not=0$
and $p=p(\mu)$) and therefore the $C_3$ is not maximal. Also, without making
any assumptions on the equation of state, it can be easily seen that either the
$C_3$ is not maximal or $X$ is not a proper CKV. \hfill\break

Type $III$: The only solution of this type which admits no further CKVs
(including KVs and HVFs) is
\be
ds^2=e^{2f}\left\{ -dt^2+dr^2+{e^{-2t}\over (\cosh r)^{2(A+1)}}dy^2+
{\sinh^2r\over (\cosh r)^{2(A+1)}}dz^2\right\}\ ,\label{c36}
\ee
where
\be
f=-{A\over 2}U -\ln(\cosh kU) +A\ln(\cosh r)+ \alpha\ ,
\ee
 $A$ and $\alpha$ are constants, $U=U(t,r)$ is given by
\be
U=t+\ln(\cosh r)\ ,
\ee
the constant $k$ takes the value,
\be
k={\sqrt{(A+1)^2+1}\over 2}\ ,
\ee
and $\mu$ and $p$ are given by,
\bea
\mu e^{2f}&=&{1\over \cosh^2r}\{1+A+3(M-1)^2\}\ , \\
p e^{2f}&=&{1\over \cosh^2r}\{-1+2A-3M^2+ (4-2A)M\}\ ,
\eea
where $M=M(U)$ is defined by
\be
M\equiv - {A\over 2}-k\tanh(kU)\ .\label{c42}
\ee
The imposition of the energy conditions ($\mu>0$, $-\mu\leq p\leq \mu$)
results in
restrictions on the values that the parameter $A$ can take, namely $
A\in  (-1,5)$. \hfill\break

Type $V$: In this case $X$ cannot be a proper CKV. However, there are
solutions belonging to this type for which $X$ is a HVF (see for instance
\cite{Ali1}). \hfill\break

Type $VI$: No perfect fluid solutions of this type exist which admit a
proper CKV, as can  easily be seen from the field equations. \hfill\break

{\bf Family B:}\hfill\break

All the solutions in this family which satisfy the conditions set up at the
beginning of this subsection are either  such that $X$ becomes a KV (metrics of
the form (\ref{c27})), or they admit two further KV which together with $\xi$
and $\eta$ generate a four parameter group of isometries $G_4$ acting on
three-dimensional spacelike orbits, the space-time thus becoming spatially
homogeneous (metrics of the form (\ref{c25}) and (\ref{c26})). Furthermore,
the subgroup $G_3$ that $G_4$ necessarily contains acts on two-dimensional
spacelike orbits, coordinated by $r$ and $z$, which are then of positive
constant curvature; hence the models are spherically symmetric Kantowski-Sachs
space-times.

\subsection{Fluid flow orthogonal to the conformal orbits.}

The conformal orbits must be spacelike in this case, and the coordinates are
comoving, i.e.
\be
u_t=-e^F \quad u_r=0\ .
\ee
 \hfill\break

{\bf Family A:}\hfill\break

Type $I$: The only solutions belonging to this type which admit no further
symmetry are
\be
ds^2={f_o^2\over \cosh^2r}\left\{-dt^2+dr^2+{(\cosh t)^{\sigma+1}\over
(\sinh t)^{\sigma-1}}dy^2 +{(\sinh t)^{\sigma+1}\over
(\cosh t)^{\sigma-1}}dz^2\right\}, \label{c44}\ee
where $f_o$ is an arbitrary  constant and the parameter $\sigma$ must be such
that $\vert \sigma\vert<1$; the energy density and pressure are then given by
\bea
\mu&=& {1\over f_o^2}\left\{(1-\sigma^2){\cosh^2 r\over\sinh^2
2t}+3\right\},\label{c45} \\
p &=& {1\over f_o^2}\left\{(1-\sigma^2){\cosh^2
r\over\sinh^2 2t}-3\right\}.\label{c46}  \eea
The equation of state  is simply $p=\mu-6/f_o^2$ as can be easily seen
from  (\ref{c45}) and (\ref{c46}). Notice that this solution is only valid in
the region $t>0$ and that it is separable in the variables $t$ and $r$; and
hence it must be  contained in the solutions given by Ruiz and
Senovilla \cite{Ruiz}.\hfill\break

Type $III$: No perfect fluid solutions admitting a proper CKV exist,
since $\mu+p=0$. \hfill\break

Type $V$: There are no solutions of this type for perfect fluids with $\mu>0$
and the
conditions set up above, as can  easily be seen from the field equations.
\hfill\break

Type $VI$: In this case, the line element, density and pressure take the
following forms:
\be
ds^2=e^{2f(t,r)}\left\{-dt^2+dr^2+Q(t)\left[{e^{-2r}\over h^2(t)}dy^2+
e^{-2qr}  h^2(t)dz^2\right] \right\},\label{c51}
 \ee
\be
\mu e^{2f}=(\dot\beta^2-1)\left\{ {3m^2\over (1+e^{-mu})^2}+qn^2{1\over
\dot\beta^2}\right\},
\ee
\be
p e^{2f}=(\dot\beta^2-1)\left\{ qn^2{1\over
\dot\beta^2}-{3m^2\over (1+e^{-mu})^2}+4q{1+q^2\over(1+q)^2}{1\over 1+e^{-mu}
}\right\},
 \ee
where $q\not=0,1$ and
\be
m\equiv{1+q^2\over 1+q} \quad n\equiv{1-q\over 1+q},
\ee
and
\be
f=H(u)+\lambda(t) \quad u\equiv r+\beta(t),
\ee
where $H(u)$ is given by
\be
H=-\ln(1+e^{-mu}),\label{c56}
\ee
and there are two possibilities for $\beta(t)$, namely: \hfill\break

\be
({\rm i}) \quad \beta=-{1\over k}\ln\vert \sinh kt\vert,\label{c57}
\ee
with $k\equiv -{(q-1)^2\over q+1}$, whence
\be
Q=\cosh kt \quad h^2=(\cosh kt)^{1/n},
\ee
\be
\lambda=-{1+q^2\over(1-q)^2}\ln\vert \sinh kt\vert+\lambda_o\ , \quad
\lambda_o=const. \label{c59}
\ee

\be
({\rm ii}) \quad \beta=-{1\over k}\ln( \cosh kt), \label{c60}
\ee
whence
\be
Q=\sinh kt \quad h^2=(\sinh kt)^{1/n},
\ee
\be
\lambda=-{1+q^2\over(1-q)^2}\ln( \cosh kt)+\lambda_o\ , \quad
\lambda_o=const.\label{c62}
\ee
Again, these solutions are only valid for $t>0$. It is also easy to see that
the choice of a negative value for the parameter $q$ would imply that
$(\mu+p)$ and $(\mu-p)$  have opposite signs, thus violating one of the
energy conditions (furthermore, $\mu$ could only be non-negative in a certain
region of the space-time). On the other hand, assuming $q>0$ immediately
leads to the choice (i) for $\beta(t)$ in order to have $\mu>0$, whence $\mu+p$
is always non-negative but there will always be some region of the space-time
in which $\mu-p<0$; consequently these solutions are only valid in a certain
region of the space-time (in particular: $(t,r)$ satisfying $\sinh \vert k\vert
t< [(q^2-2q/3+1)/2q]^{k/m}e^{kr}$). \hfill\break

{\bf Family B:}\hfill\break

It is easy to see from the field equations that there are no perfect fluid
solutions in this family.

\section{Perfect fluid space-times: Tilted case.}
\label{se5}
The field equations in this case are (\ref{c32}) and (\ref{c33}), but we no
longer have the additional condition $u_r=0$ as in the previous cases. However,
it
should be noted  that for a perfect fluid solution admitting an
Abelian $G_2$ of isometries acting on spacelike orbits and such that they
admit orthogonal surfaces, it is always possible to perform a change of
coordinates so as to bring the coordinates into a comoving form with respect
to $u^a$, while leaving
the metric diagonal \cite{Wain}. Such a coordinate change in the $t,r$ plane
(in our coordinates) would dramatically change the form of the proper CKV $X$;
for example, we would no longer be able to integrate out the conformal
equations (\ref{c3}) nor  provide simple expressions for the metric
functions. Roughly speaking, one must choose between the coordinate chart
adapted to the CKV and the one adapted to the four-velocity of the fluid.
It is
interesting to notice that this is not the case for HVFs; since one
may change there to comoving coordinates  the HVF changing then in a \lq\lq
controlled" way, so that one can still  integrate out the homothetic equations
(see for instance \cite{Ali1}). This difference is mainly due to the fact that
the four-velocity field of a perfect fluid and a HVF are always surface forming
(or
in a more physical language, the fluid \lq\lq inherits" the symmetry, see
\cite{ColTup90a}), whereas this is not necessarily so in the case of a CKV.

We shall deal separately with the cases of timelike and spacelike conformal
orbits and,         in each case, we shall distinguish between the two
families of
metrics A and B.

\subsection{Timelike conformal orbits.}
\label{se51}
\indent{\bf Family A:}\hfill\break

We shall discuss here some general features of the solutions belonging to
different types in this family. \hfill\break

The metric can be  written as:
\be
ds^2=e^{2F(t,r)}\left\{-dt^2+dr^2+e^{at}{q(r)\over h^2(r)}dy^2+
e^{bt}q(r)h^2(r)dz^2\right\}, \label{cc71} \ee
where
\begin{center}
\begin{tabular}{|c|c|c|} \hline
Type & $a$ & $b$ \\ \hline
$I$  & 0    & 0  \\ \hline
$III$  & -2  & 0  \\ \hline
$V$  & -2 & -2  \\ \hline
$VI$  & -2    & -2q  \\ \hline
\end{tabular}
\end{center}
and $q\not=0,1$.

For these cases Eqn. (\ref{c32}) gives:
\be
{a^2-b^2\over 4}-2\left( {h'\over h}\right)^2+2{h'q'\over hq} +2{h''\over h}+
4{h'\over h}F_{,r}+(a-b)F_{,t}=0\ ,
\label{cc72}
\ee
where a dash indicates differentiation with respect to $r$. At this point we
must distinguish between two cases depending on whether $a=b$ or not.
\hfill\break

\underline{Case (i):} Assume $a=b$ (types $I$ and $V$). Equation (\ref{cc72})
reads now:
\be
0=\left( {h'\over h}\right)'+{h'q'\over hq} + 2{h'\over h}F_{,r}
\label{cc73}
\ee
since $h'\not=0$ (otherwise the metric admits a further KV), this equation
immediately gives
\be
e^{2F}=f(t){h\over q h'}
\label{cc74}
\ee

\underline{Case (ii):} If $a\not=b$ (types $III$ and $VI$). Differentiating
(\ref{cc72}) with respect to $t$, we have
\be
0=4{h'\over h}F_{,rt}+(a-b)F_{,tt}\ ,
\label{cc75}
\ee
that integrates to give
\be
F=H(t+\beta (r)) +\lambda (r)\ ,
\label{cc76}
\ee
with
\be
\beta'{h'\over h}={b-a\over 4}\ ,
\label{cc77}
\ee
and the whole equation (\ref{cc72}) reads
\be
{a^2-b^2\over 4}+2\left( {h'\over h}\right)'+2{h'q'\over hq}+
4{h'\over h}\lambda'=0\ .
\label{cc78}
\ee

Let us deal next with the first case. For types $I$ and $V$ we can use an
alternative form of the metric that satisfies (\ref{c32}) identically, in
order to simplify the resulting expressions:
\be
ds^2={e^{f(t)-q(r)}\over h'(r)}\left\{-dt^2+dr^2\right\}+ {e^{f(t)+at}\over
h'(r)}\left\{ e^{-h(r)} dy^2 +  e^{h(r)} dz^2 \right\},
\label{cc79}
\ee
the conformal factor now being:
\be
\Psi = {f_{,t}\over 2}\ .
\ee
Equation (\ref{c33}) can be written as
\be
0=\Sigma_0 +a\Sigma_1 f_{,t} +\Sigma_2 f_{,t}^2 +{a\over 4} f_{,t}^3 +\Sigma_3
f_{,tt}-{a\over 2}f_{,t}f_{,tt}\ ,
\label{cc80}
\ee
where $\Sigma_i$, $i=0,\ldots,3$,  denote functions depending only on $r$
\bea
\Sigma_0 &=& {a^2\over 4}(q')^2 + {a^2\over 4}{h''\over h'}q' +{1\over 4}h''h'q'
-{1\over 4}\left( {h''\over h'}\right)^2(q')^2 +{1\over 4}\left( {h''\over
h'}\right)^3q'-{a^2\over 4}q'' \nonumber \\
& & -{1\over 4}(h')^2q''-{1\over 4}\left( {h''\over h'}\right)^2q'' +{1\over
4}(q'')^2 -{1\over 2}{h'''h''\over (h')^2}q' +{1\over 2}{h'''\over h'}q''\
,\label{z70} \\
\Sigma_1 &=&{a^2\over 4} +{1\over 4}(h')^2+{1\over 2}(q')^2 +
{1\over 2}{h''\over h'}q' + {1\over 4}\left( {h''\over h'}\right)^2  -{1\over
2}q'' -{1\over 2}{h'''\over h'}\ ,\\
\Sigma_2 &=&{a^2\over 2} +{1\over 4}(h')^2+{1\over 4}(q')^2 +
{1\over 4}{h''\over h'}q' + {1\over 2}\left( {h''\over h'}\right)^2  -{1\over
4}q'' -{1\over 2}{h'''\over h'}\ ,\label{z72}\\
\Sigma_3 &=&-{a^2\over 2} -{1\over 2}(h')^2+
{1\over 2}{h''\over h'}q' - {1\over 2}\left( {h''\over h'}\right)^2  +{1\over
2}q'' +{h'''\over h'}\ ,\label{z73}
\eea
Since $f_{,tt}\not=0$ (otherwise $X$ would become a HVF) the following
possibilities arise from Eqn. (\ref{cc80}) namely: \hfill\break

($I.a$) $\Sigma_3=0$ then $\Sigma_2=\Sigma_0=0$ (for all $f(t)$ ).\hfill\break

($I.b$) $\Sigma_3\not=0$ then $\Sigma_2=-c\Sigma_3$, $\Sigma_0=\alpha\Sigma_3$,
\be
0=\alpha -c(f_{,t})^2 + f_{tt}
\label{cc81}
\ee
where $\alpha$  and $c$ are constants. \hfill\break

($V.a$) $\Sigma_i=\alpha_i$ (constants). \hfill\break

($V.b$)
\be
0=e+d f_{,t}+{1\over 2}(f_{,t})^2- f_{,tt}
\ee
\be
0=2\Sigma_2+\Sigma_3+2d,
\ee
\be
0=\Sigma_1+d\Sigma_3 +e,
\ee
\be
0=\Sigma_0+e\Sigma_3,
\ee
where $e$, $d$ are constants.

For all these cases one has:
\bea
G_{tr}&=&-{a\over 2}q' -{1\over 2}f_{,t}\left(q'+{h''\over h'} \right)\ ,\\
\mu g_{rr}&=& {3\over 4}a^2+{3\over  2}af_{,t} +{3\over 4}(f_{,t})^2 +{1\over
4}(h')^2-{3\over 4}\left( {h''\over h'}\right)^2 -{1\over 2}q''\ ,\\
(\mu-p)g_{rr}&=& a^2+2af_{,t}+(f_{,t})^2+ f_{,tt}-2\left( {h''\over h'}\right)^2
+{h'''\over h'}\ ,\\
(\mu+p)g_{rr}&=& {a^2\over 2}+af_{,t}+{1\over 2}(f_{,t})^2 -f_{,tt}+{1\over
2}(h')^2+{1\over 2}\left( {h''\over h'}\right)^2- q'' -{h'''\over h'}\  ,\\
(\mu+p)u_t^2&=& {a\over 2}f_{,t}+{1\over 2}(f_{,t})^2 -f_{,tt} +{1\over
2}{h''\over h'}q'-{1\over 2}q'' \ .\label{z83}
\eea

\underline{Subcase $I.a$}  $a=0$ and
$\Sigma_0=\Sigma_2=\Sigma_3=0$. It can be trivially  checked that
\be
q'+{h''\over h'}=0,
\ee
and that $G_{tr}=0$, so the solution has a comoving fluid flow and, as  was
pointed out before, it has more symmetry. \hfill\break

\underline{Subcase $I.b$}  $\Sigma_2=-c\Sigma_3$ yields
\be
0=(1-2c)(h')^2+(q')^2 +(1+2c){h''\over h'}q'+2(1-c)\left( {h''\over h'}\right)^2
-(1-2c)q''-2(1-2c){h'''\over h'}.
\label{cc82}
\ee
For the particular case $c={1\over 2}$ the above equation reads
\be
\left(q'+{h''\over h'}\right)^2=0 \ ,
\ee
and then $\Sigma_0=\alpha\Sigma_3$ implies
\be
{1\over 2}{h''' \over h'}-\left( {h''\over h'}\right)^2=-\alpha \ ,
\ee
and one can see from Eqn. (\ref{z83}) that $(\mu+p)u_t^2=0$, so, this solution
cannot represent a perfect fluid. For
$c\not={1\over 2}$, we obtain, after some calculations involving (\ref{cc82})
\bea
\Sigma_0&=&{1\over 4(1-2c)}\left[q''-q'{h''\over h'} \right] \left[
q'+{h''\over h'} \right]^2 \ ,\\
\Sigma_3&=&{1\over 2( 1-2c)}\left[
q'+{h''\over h'} \right]^2 \ .
\eea
The remaining equation $\Sigma_0=\alpha\Sigma_3$ can be written as
\be
q''-q'{h''\over h'}=2\alpha\ .
\label{cc83}
\ee
Defining  now
\be
\sigma'\equiv{1\over h'} \ ,\label{c2c83}
\ee
equation (\ref{cc83}) can be integrated and (\ref{cc82}) rewritten to give
\be
q'=2\alpha {\sigma\over\sigma'}+{k\over \sigma'} \ , \quad k={\rm const} \ ,
\ee
\be
0=\sigma'\sigma'''+{-1+3c\over 1-2c}(\sigma'')^2-{2c\over
1-2c}\sigma''(2\alpha\sigma +k)-\alpha(\sigma')^2+{1\over
2(1-2c)}(2\alpha\sigma +k)^2 +{1\over 2}\ , \label{cc84}
\ee
and one then has
\bea
(\mu+p)u_t^2&=& {1-2c\over 2}(f_{,t})^2 \ , \\
(\mu+p)g_{rr}&=&{1-2c\over 2}(f_{,t})^2 -{1\over 2(1-2c)} \left[ 2\alpha
{\sigma\over\sigma'}+{k\over \sigma'}-{\sigma''\over \sigma'}\right]^2\ ,
\label{cc85}\\
(\mu-p)g_{rr}&=& (1+c)(f_{,t})^2-\alpha-{\sigma'''\over \sigma'}\ .
\label{cc86}
\eea
In order to satisfy the energy conditions in all regions of the space-time, the
right-hand sides of the above equations must all be positive; necessary
conditions  for this are $c<{1\over 2}$ and also that $(f_{,t})^2$ has a
strictly  lower bound. The latter condition singles out one of the
possible solutions to (\ref{cc81}) namely:
\be
\alpha=c\beta^2 \quad f=-{1\over c}\ln \vert \sinh (c\beta t+C_1
)\vert\ .\label{cc87} \ee
where another additive constant of integration in the expression of $f$ has
been set equal to zero without loss of generality. Notice that $C_1$ can also
be made zero by suitably redefining $t$, nevertheless we choose to  maintain
it in order to allow singularity-free solutions when $t=0$.

One particular solution to (\ref{cc84}) is
\be
\sigma={\cosh \beta r \over \beta^2} \quad {\rm and}\quad k=1-2c\ . \label{cc88}
\ee
 Notice that in order to have the correct
signature in the metric $r$ is restricted to positive values only. The energy
condition (\ref{cc86}) is satisfied for all values of $r$ but only if
$c>-1$. \hfill\break

Type $V$: After some calculations, it can be seen that the only possibility for
the line element given by (\ref{cc79}) leading to an Einstein tensor of the
perfect fluid type is:
\be
ds^2=e^{f(t)-2ar}\{-dt^2+dr^2\}+e^{f(t)-2t}\{e^{-mr}dy^2+e^{mr}dz^2\}
\label{c94} \ee
where $a$ and $m$ are constants. The only remaining  field equation
is then
\be
0=4a^2-(2+4a^2+{m^2\over 2})f_{,t} +(2+a^2+{m^2\over 4})(f_{,t})^2- {1\over
2}(f_{,t})^3 +f_{,tt}(f_{,t}-2-{m^2\over 2}),
\ee
that integrates to give
\be
ke^t=(f_{,t}-2)(f_{,t}-p^2(1+s))^{(1-s)/2s}(f_{,t}-p^2(1-s))^{-(1+s)/2s},
\ee
where the constants $p^2$ and $s$ are defined in terms of $a$ and $m$ by
\be
p^2=a^2+{m^2\over 4}+1 \ ,\quad s^2=1-{4a^2\over p^4}\ .
\ee
The density, pressure and velocity field are given by
\bea
(\mu+p)u_t^2&= & {a^2 (f_{,t}-2)^2 \over f_{,t}-2(p^2-a^2) }\ ,\label{c95}\\
(\mu+p)g_{rr}& =& -( f_{,t}-2(p^2-a^2))+  {a^2 (f_{,t}-2)^2 \over
f_{,t}-2(p^2-a^2) }\ ,\\
(\mu-p)g_{rr}&=& { (f_{,t}-2)({3\over
2}(f_{,t})^2+(-3p^2+2a^2-2)f_{,t}+4p^2-2a^2) \over f_{,t}-2(p^2-a^2) }\ .
\label{c98}
 \eea
The energy conditions $\mu\pm p\ge 0$ will be satisfied if and
only if the right-hand sides of the above equations are positive, i.e.:
\bea
& & f_{,t}-2(p^2-a^2)>0 \ , \\
& &(a^2-1)(f_{,t})^2+ (4p^2-8a^2)f_{,t} +4a^2-4(p^2-a^2)^2>0 \ , \\
& &{3\over
2}(f_{,t})^2+(-3p^2+2a^2-2)f_{,t}+4p^2-2a^2\ge 0 \ ,
\eea
and these, in turn, can be seen to imply (after a careful analysis):
 \bea
a^2-1\ge 0 &\qquad &f_{,t}>\beta   \\
a^2-1<0 &\qquad & \beta<f_{,t}<2\left(1+{m^2\over 4}{1\over 1-\vert a\vert} 
\right)  \eea
where
\be
\beta\equiv {1\over
3}\left[ a^2+3{m^2\over 4}+5+\sqrt{(a^2-1)^2+3{m^2\over 4} (2+2a^2+3{m^2\over
4})} \right] .
 \ee

From (\ref{c95})-(\ref{c98}) it is easy to see that a barotropic equation
of state  of the form $p=p(\mu)$ is not possible.\hfill\break

The analysis of types $III$ and $VI$ turns out to be very involved, with many
subcases arising, and so far, we have not been able to find any solution to
the EFEs which is valid (i.e., satisfies energy conditions) all over the
space-time manifold. Since we cannot provide any result in the positive we
shall not discuss these cases any further here, so as to keep the present
study at a reasonabe length. \hfill\break

{\bf Family B:}\hfill\break

For this family, the metric and CKV $X$ take  one of the forms given by
(\ref{c19})-(\ref{c21}), and the conformal factor $\Psi$ is then given by
(\ref{c22}).

From the field equations for a perfect fluid it is easy to see that the
metrics
(\ref{c20}) and (\ref{c21}) cannot represent a perfect fluid space-time with
$(\mu+p)\not=0$ and $\mu>0$; thus, only (\ref{c19}) needs be considered. The
field equations (\ref{c32})-(\ref{c33}) for this metric suggest
redefinitions of the coordinates $t=\phi_1(t)$  and $r=\phi_2(r)$, so that
the
metric takes the form (in the new coordinates):
\be
ds^2={e^{2f(t-r)}\over \sqrt{H'}}\left\{-{dt^2\over t}+{H'\over H}dr^2
+4tdy^2+Hdz^2\right\}\ ,\label{c101}
 \ee
where $H=H(r)$, as usual a dash indicates a derivative with respect to $r$ and
$f=f(t-r)$ is a function of $t-r\equiv x$. The CKV $X$ and conformal factor
$\Psi$ are in these coordinates:
\be
X=e^y\left( t^{1/2}\partial_t -{1\over 2}t^{-1/2}\partial_y \right)\ ,
\ee
\be
\Psi=e^y\sqrt{t}f_{,x}\ .\label{c103}
\ee

The remaining field equations are now:
\be
(f_{,x}^3-f_{,x}f_{,xx})\Sigma_0(r)+ f_{,xx}\Sigma_1(r)
+f_{,x}^2\Sigma_2(r)=0\ , \label{c104} \ee
where
\be
\Sigma_0(r)\equiv 2 {H''\over H'}\ , \quad
\Sigma_1(r)\equiv -{5\over 4}\left({H''\over H'}\right)^2 +{H'''\over H'}  \ ,
\ee
\be
\Sigma_2(r)\equiv {3\over 2}\left({H''\over H'}\right)^2 -{H'''\over H'}  \ ,
\ee
and they satisfy: $\Sigma_1+\Sigma_2={\Sigma_0^2\over 16}$.

Also notice that one of the trivial solutions to (\ref{c104}), namely
$f_{,x}=0$ would correspond to $X$ being a KV, on account of the form of the
conformal factor (\ref{c103}) and that the case $f_{,x}=$constant leads to an
incorrect signature of the metric.
 Excluding these cases, $\Sigma_0$, $\Sigma_1$
and $\Sigma_2$ must be constants, and the following possibilities then
arise:\hfill\break

(B-1) $\quad \Sigma_0=0$, then necessarily $\Sigma_1=\Sigma_2=0$, and
\be
H=K^2r+C\ ,\quad K,C={\rm const}\ ,
\ee
where we can always make $C=0$ by suitably redefining $r$, and $f$ is then a
completely arbitrary function of its argument. The density, pressure and
velocity
field can then be obtained from:
\bea
\mu {g_{yy}\over 4t} & =& 3\{ f_{,x}+(t-r)f_{,x}^2\}, \\
(\mu+p) {g_{yy}\over 4t} & =& 2(t-r)\{ f_{,x}^2-f_{,xx} \} , \\
(\mu-p) {g_{yy}\over 4t} & =& 2\{(t-r)f_{,xx}+2(t-r)f_{,x}^2+3f_{,x} \}, \\
(\mu+p)u_t^2 & =& 2\{f_{,x}^2-f_{,xx} \} .
\eea
From the above expressions one can readily see that no function $f(x)$
exists such that $\mu$, $\mu+p$ and $\mu-p$ are positive over all of the
space-time. \hfill\break

(B-2) $\quad \Sigma_0\not =0$. This leads to
\be
H=K^2\{e^{4sr}+h_0\}\ ,
\ee
where $K$, $s$ and $h_0$ are arbitrary constants. The function $f$ is then
given implicitly  by
\be
e^{s/f_{,x}}\left\{ 1+{s\over f_{,x}}\right\}=e^{-2sx}\ ,\label{c118}
\ee
and the density and pressure can be obtained from:
\bea
\mu {g_{yy}\over 4t} & =&{3e^{-4sr}\over 4s}\left\{
e^{4sr}\left[f_{,x}^2(4st-1) +2sf_{,x} +{s^2\over 3} \right]-h_0(s+f_{,x})^2
\right\}\ , \\
 (\mu+p){g_{yy}\over 4t} & =&{e^{-4sr}\over 2(2f_{,x}+s)}\left\{
e^{4sr} \left[-f_{,x}^2(4st-1) +2sf_{,x} +s^2 \right] \right.\nonumber\\
\, & + & \left. h_0(s+f_{,x})^2 \right\}\ , \\
(\mu-p){g_{yy}\over 4t} & =&{e^{-4sr}\over s(2f_{,x}+s)}\left\{ 3
e^{4sr}f_{,x} \left[f_{,x}^2(4st-1) +{4s\over 3}(2st+1)f_{,x} +s^2 \right]
\right.\nonumber\\ \, & - & \left. h_0\left[ 3 f_{,x}^3+8sf_{,x}^2
+7s^2f_{,x}+2s^3 \right] \right\}\ .
\eea
The energy conditions $\mu>0$, $\mu\pm p>0$ will be satisfied if and only if
the above expressions are all  positive, and a careful  analysis of these
conditions taking into account (\ref{c118}) shows that they can  only  hold in
some open domains of the space-time manifold (e.g., for $h_0=0$ the energy
conditions can only hold in the region $t-r<0$ if $s<0$, and in the region
$-{\cal K}^2(t)<t-r<0$ if $s>0$, ${\cal K}$ being some function  of  $t$).

\subsection{Spacelike conformal orbits.}
\label{se52}

\indent{\bf Family A:}\hfill\break

The forms of the metric functions are then (\ref{c11}), (\ref{c13}),
(\ref{c15}) and (\ref{c16}), with   $w=0$ and the coordinates $t$ and $r$
interchanged. We note that the solutions are not simply obtained from
(\ref{se51}) by interchanging $t$ and $r$. We next briefly discuss a few
solutions belonging to some of the Bianchi types.\hfill\break

Type $I$: The analysis of this type follows much along the same lines as its
counterpart in the case of timelike conformal orbits. Thus we have
\be
ds^2={e^{f(r)-q(t)}\over \dot h(t)}\left\{-dt^2+dr^2\right\}+ {e^{f(r)}\over
\dot h(t)}\left\{ e^{-h(t)} dy^2 +  e^{h(t)} dz^2 \right\}\ ,
\label{cc130}
\ee
where a dot indicates differentiation with respect to $t$. Then the
independent Einstein equation takes the form:
 \be
 0=\hat\Sigma_0+\hat\Sigma_2(f_{,r})^2+\hat\Sigma_3 f_{,rr}\ ,
 \label{cc131}
 \ee
where the expressions of the $\hat\Sigma_i$s can be formally obtained from
those of the $\Sigma_i$s given by Eqns. (\ref{z70}), (\ref{z72}) and
(\ref{z73}), by simply changing $r$ by $t$ there (i.e., changing primes into
dots).

Since $f_{,rr}$ is non-null, the following possibilities arise: \hfill\break

($I.a$) $\hat\Sigma_3=0$, then necessarily $\hat\Sigma_0=\hat\Sigma_1=0$.
In this
subcase there are no perfect fluid solutions because the Einstein tensor do not
have any timelike eigenvector.\hfill\break

($I.b$) $\hat\Sigma_3\not=0$, then $\hat\Sigma_2=-c\hat\Sigma_3$,
$\hat\Sigma_0=\alpha\hat\Sigma_3$ and
\be
f_{,rr}=c(f_{,r})^2-\alpha\ ,
\label{z132}
\ee
where $c$ and $\alpha$ are arbitrary constants. The first equation yields
  \be
0=(1-2c)(\dot h)^2+(\dot q)^2 +(1+2c){\ddot h\over \dot h}\dot q+2(1-c)\left(
{\ddot h\over \dot h}\right)^2 -(1-2c)\ddot q-2(1-2c)
{{h}^{\!\cdots}\over\dot h}
\label{cc132}
\ee
The case $c={1\over 2}$  can be easily  seen to correspond to one of the
previously studied cases, namely those  with the fluid flow
orthogonal to the  conformal orbits.

Following now a similar procedure to the one outlined in the case of timelike
conformal orbits, the field equations can then be written as

\be
\dot q=2\alpha {\sigma\over\dot\sigma}+{k\over \dot\sigma}\ , \quad k={\rm
const},
\ee
\be
0=\dot\sigma {\sigma}^{\!\!\!\!\!\cdots}+{-1+3c\over
1-2c}(\ddot\sigma)^2-{2c\over 1-2c}\ddot\sigma(2\alpha\sigma
+k)-\alpha(\dot\sigma)^2+{1\over 2(1-2c)}(2\alpha\sigma +k)^2 +{1\over 2}\ ,
\label{cc134}
\ee
with $\sigma(t)$ defined through
\be
\dot\sigma={1\over \dot h}\ .
\ee
Explicit expressions for $\mu$, $p$ and $u_t$ can be readily derived from:
\bea
(\mu+p)u_t^2&=&{1\over 2(1-2c)} \left[ 2\alpha
{\sigma\over\dot\sigma}+{k\over \dot\sigma}-{\ddot\sigma\over
\dot\sigma}\right]^2\ ,\\
(\mu-p)g_{rr}&=&-(1+c)(f_{,r})^2 +\alpha +{{\sigma}^{\!\!\!\!\!\cdots}  \over
\dot \sigma}\ , \\
(\mu+p)g_{rr}&=&{1\over 2(1-2c)} \left[ 2\alpha
{\sigma\over\dot\sigma}+{k\over \dot\sigma}-{\ddot\sigma\over
\dot\sigma}\right]^2 -{1-2c\over 2}(f_{,r})^2\  .
\label{cc136}
\eea

Now, from all the possible solutions for $f(r)$ to equation (\ref{z132}),
 the energy conditions $\mu\pm p>0$ single out
\be
\alpha=c\beta^2 \ , \quad f=-{1\over c} \ln (\cosh c\beta r)
\ee
and restrict $c$ to values $c<1/2$.

A particular solution to (\ref{cc134}) which satisfies the energy conditions is:
\be
\sigma={\cosh \beta t \over \beta^2} \ ,\quad k=-(1-2c)\ , \label{cc138}
\ee
and we notice that the solution is valid for $c>-1$ and in order for the metric
to have the correct signature, $t$ must be positive.\hfill\break

Type $V$: There are no perfect fluid solutions of this type.\hfill\break

{\bf Family B:}\hfill\break

An inspection  of the field equations suggests, as in the case of timelike
conformal orbits, a redefinition of $t$ and $r$, so that in the new
coordinates the metric, CKV $X$ and the conformal factor $\Psi$ take the
following forms:
\be
ds^2={2 e^{2f(x)} e^{-kt/2}\over \sqrt{\dot H}}\left\{-{1\over 4}{\dot H\over
H}dt^2 + {k\over 4}{ce^{kr}\over 1+ce^{kr}}dr^2+{1+ce^{kr}\over k}dy^2
+Hdz^2\right\}\ , \label{MM}
 \ee
\be
X=e^y\left\{e^{-kr/2}\sqrt{1+ce^{kr}}\partial_r -{k\over 2}{ce^{kr/2} \over
\sqrt{1+ce^{kr}} }\partial_y\right\}\ ,
 \ee
\be
\Psi=-e^yf_{,x}e^{-kr/2}\sqrt{1+ce^{kr}}\ ,\label{c137}
\ee
where $k$  and $c$ are constants and $x$ is defined as $x\equiv t-r$. The
field equations
then reduce to:
\be
2(f_{,x}f_{,xx}-f_{,x}^3)\Sigma_0 + f_{,x}^2\Sigma_1 + (f_{,xx}+{k\over
2}f_{,x})\Sigma_2=0 \ , \label{c138}
\ee
where $\Sigma_0(t)$, $\Sigma_1(t)$ and $\Sigma_2(t)$ are given by:
\be
\Sigma_0\equiv -k+{\ddot H \over \dot H} \ ,
\ee
\be
\Sigma_1\equiv -k^2+{k\over 2}{\ddot H\over \dot H}+{1\over 2}\left( {\ddot
H\over \dot H} \right)^2-\left( {\ddot H\over\dot H} \right)^{\cdot} =
-\dot\Sigma_0 +{1\over 2}{\Sigma_0}^2 +{3\over 2}k\Sigma_0 \ ,
\ee
\be
\Sigma_2\equiv {k^2\over 4}-{1\over 4}\left( {\ddot
H\over \dot H} \right)^2 +\left( {\ddot H\over\dot H} \right)^{\cdot} =
\dot\Sigma_0-{1\over 4}\Sigma_0(\Sigma_0+2k) \ .
\ee
The possibility $f_{,x}=const\not=0$ leads to the wrong Segre type of the
Einstein tensor (i.e., no perfect fluid solutions exist). Then,
from Eqn.(\ref{c138}) it is immediate  that either $f_{,x}=0$ (and then
$X$ becomes a KV; see (\ref{c137}) ) or $\Sigma_0$   and $\Sigma_1$ are both
constants, which, in turn, implies:

\be
H=Me^{dt}+m\ ,
\ee
where $M$, $m$ and $d$ are constants. Substituting this back into the field
equation (\ref{c138}) we obtain:
\be
-2(d^2-k^2)f_{,xx}+16(d-k)(f_{,x}f_{,xx}-f_{,x}^3)+4(d^2-k^2+k(d-k))f_{,x}^2
-k(d^2-k^2)f_{,x}=0,\label{c142}
 \ee
and we  can distinguish the following cases:\hfill\break

(i) $\quad f_{,xx}=0$; i.e., $f_{,x}=$constant,   and from Eqn.(\ref{c142})
we obtain
\be
f_{,x}={k+d\over 4} \quad {\rm or} \quad f_{,x}={k\over 4}.
\ee
In the first case ($f_{,x}={k+d\over 4}$)  the Segre type does not correspond
to a perfect fluid (since $u_t^2=0$) and in the second case  $(\mu+p)u_t^2<0$,
so this cannot represent a physical perfect fluid. \hfill\break

(ii) $\quad f_{,xx}\not=0$; whence Eqn.(\ref{c142}) can be integrated once,
leading to an implicit expression for $f_{,x}$. A careful analysis of the
energy
conditions $\mu\pm p\ge 0$ shows that in the general case (i.e., no
assumptions on the values of the parameters $k$ and $d$), they {\it can
only} be
satisfied over certain restricted open domains of the space-time. However,
two special cases arise  which deserve special attention;
namely, $k=d$ and $k=-d$. For $k=d$, Eqn.(\ref{c142}) is identically
satisfied,
$f$ thus being a completely arbitrary function; however, this
does not correspond to a perfect fluid (it is of the wrong Segre type).
In the second
case, Eqn.(\ref{c142}) can be integrated  to give:
\be
\label{efe}
f=\ln{e^{kx/4}\over{4\over k}+ae^{kx/4}},
\ee
$a$ being a constant, and $\mu$, $p$   and $u_t$ can then be obtained from:
\bea
(\mu+p)&=& {e^{-kx/2}\over c\sqrt{-kM}}(4+ak\, e^{kx/4})(-cm\,
e^{kt}-Me^{-kr}-2cM) \ , \\
(\mu-p)&=& {e^{-kx/2}\over c\sqrt{-kM}}\left[ (16+ak\,
e^{kx/4})(-cm\, e^{kt}-Me^{-kr}) \right. \nonumber \\
& \, & \left. + ackM\,
e^{kx/4}(2-ak\, e^{kx/4} ) \right] \ , \\ (\mu+p)u_t^2 &=&{-k^2\over
2(4+ake^{kx/4}) }\ . \eea
As usual, the energy conditions will be fulfilled if and only if the above
expressions
are all positive and, again, this can only be
possible over certain open domains of the manifold (notice that $k$ must be
positive in order for the metric to have the correct signature).

\section{Discussion}
We have studied perfect fluid space-times admitting a three-dimensional Lie
group of conformal motions containing a two-dimensional Abelian subgroup of
isometries. All such space-times have been classified geometrically and in
each
class the metric has been explicitly given in coordinates adapted to the
symmetry vectors.

In section \ref{se3} we restricted attention to the diagonal case, and in
section \ref{se4} we found all such perfect fluid solutions in which the fluid
four-velocity is tangential or orthogonal to the conformal orbits. In the
former case the orbits are necessarily timelike and the only solutions for
which ${\cal C}_3$ is maximal and $X$ is proper are of type $III$ in family A
and given by Eqns. (\ref{c36})-(\ref{c42}). In the latter case, in which $u^a$
is orthogonal to spacelike conformal orbits, there  is a 2-parameter family of
solutions given by (\ref{c44})-(\ref{c46}) of type $I$ which are valid for
$t>0$ and are separable in $t$ and $r$ (see \cite{Ruiz}); and a
class of solutions of type $VI$ [see Eqns. (\ref{c51})-(\ref{c56}) and  Eqns.
(\ref{c57})-(\ref{c59}) or (\ref{c60})-(\ref{c62})]. In this latter case,
physical constraints (e.g., the energy conditions) restrict the validity of the
solutions to a region of the space-time and, again, there are no solutions in
family B.

In section \ref{se5} perfect fluid solutions were sought in the general
(tilting) case in which the fluid four-velocity is neither tangential to nor
orthogonal to the conformal orbits. We chose to work in coordinates adapted to
the CKV and again the two cases in which the conformal orbits are timelike
(subsection \ref{se51}) or spacelike (subsection \ref{se52}) were considered.

In the timelike case, solutions in family A of types $I$ [metric (\ref{cc79}),
field eqns. (\ref{c2c83})-(\ref{cc86})] and $V$ [metric (\ref{c94}), field eqns.
(\ref{c95})-(\ref{c98})] were obtained. In both cases solutions exist such that
the energy conditions are satisfied on the whole space-time manifold; a
particular solution of type $I$ was given by Eqns. (\ref{cc87}) and
(\ref{cc88}). We noted that solutions of type $V$ cannot admit an equation
of state of the form $p=p(\mu)$. In the spacelike case, solutions in family A
of type $I$ were found and a particular solution, given by Eqn. (\ref{cc138}),
in which the energy conditions are always satisfied, was displayed. There are
no solutions of type $V$ in this case.

{\it All} solutions in family B can be found in both the timelike case
[see Eqns. (\ref{c101})-(\ref{c118})] and the spacelike case [see Eqns.
(\ref{MM})-(\ref{efe})]. However, our analysis showed that in general there
are no solutions for which the energy conditions are satisfied over the entire
space-time manifold.

The case of null conformal orbits will be studied in a future paper.

\vskip.8cm
 \noindent{\Large
\bf Acknowledgments} \vskip.8cm

The authors would like to thank M. Mars (Univ. of Barcelona) for many
helpful discussions and suggestions.
They would also like to thank the referees for many helpful suggestions  which
have greatly contributed to making the paper more precise and readable.
One of the authors (AAC) would like to
thank the Spanish Government and the \lq\lq Direcci\'{o}n  General de
Investigaci\'{o}n Cient\'{\i}fica y T\'{e}cnica (DGICYT)" for financial
assistance and  the Department of Physics at the Universitat de les Illes
Balears for its warm hospitality.   This work is partially supported by the
DGICYT Project No. PB 91-0335. Financial support from STRIDE program
(Research Project No. STRDB/C/CEN/509/92) is also acknowledged.

\end{document}